\documentclass[prx, twocolumn, superscriptaddress, floatfix, longbibliography, amsmath,amssymb, aps]{revtex4-2}

\usepackage{graphicx}
\usepackage{dcolumn}
\usepackage{bm}
\usepackage{hyperref}
\usepackage{siunitx}
\DeclareSIUnit\gauss{G}
\usepackage{braket}
\usepackage{ulem}
\usepackage[dvipsnames]{xcolor}
\usepackage{tabularx}
\usepackage{braket}

\begin{document}

\title{Observation of a Halo Trimer in an Ultracold Bose-Fermi Mixture}

\author{Alexander Y. Chuang}
\thanks{These authors contributed equally}
\affiliation{Department of Physics, MIT-Harvard Center for Ultracold Atoms, and Research Laboratory of Electronics, MIT, Cambridge, Massachusetts 02139, USA}%
\author{Huan Q. Bui}
\thanks{These authors contributed equally}
\affiliation{Department of Physics, MIT-Harvard Center for Ultracold Atoms, and Research Laboratory of Electronics, MIT, Cambridge, Massachusetts 02139, USA}%
\author{Arthur Christianen}
\thanks{These authors contributed equally}
\affiliation{Institute for Theoretical Physics, ETH Zürich, 8093 Zürich, Switzerland}%
\affiliation{%
Max-Planck-Institut für Quantenoptik, 85748 Garching, Germany
}%
\affiliation{%
Munich Center for Quantum Science and Technology (MCQST), 80799 Munich, Germany
}%
\author{Yiming Zhang}
\affiliation{Department of Physics, MIT-Harvard Center for Ultracold Atoms, and Research Laboratory of Electronics, MIT, Cambridge, Massachusetts 02139, USA}%
\author{Yiqi Ni}
\affiliation{Department of Physics, MIT-Harvard Center for Ultracold Atoms, and Research Laboratory of Electronics, MIT, Cambridge, Massachusetts 02139, USA}%

\author{Denise Ahmed-Braun}
\affiliation{Theory of Quantum and Complex Systems, Physics Department, Universiteit Antwerpen, B-2000 Antwerpen, Belgium}

\author{Carsten Robens}
\affiliation{Department of Physics, MIT-Harvard Center for Ultracold Atoms, and Research Laboratory of Electronics, MIT, Cambridge, Massachusetts 02139, USA}%

\author{Martin W. Zwierlein}
\affiliation{Department of Physics, MIT-Harvard Center for Ultracold Atoms, and Research Laboratory of Electronics, MIT, Cambridge, Massachusetts 02139, USA}%

\date{\today}

\begin{abstract}
The quantum mechanics of three interacting particles gives rise to interesting universal phenomena, such as the staircase of Efimov trimers predicted in the context of nuclear physics and observed in ultracold gases. Here, we observe a novel type of halo trimer using radiofrequency spectroscopy in an ultracold mixture of $^{23}$Na and $^{40}$K atoms. The trimers consist of two light bosons and one heavy fermion, and have the structure of a Feshbach dimer weakly bound to one additional boson. We find that the trimer peak closely follows the dimer resonance over the entire range of explored interaction strengths across an order of magnitude variation of the dimer energy, as reproduced by our theoretical analysis. The presence of this halo trimer is of direct relevance for many-body physics in ultracold mixtures and the association of ultracold molecules.
\end{abstract}
\maketitle

\section{Introduction}

Three-body bound states in quantum mechanics have a rich history, from the molecular hydrogen ion \cite{pauling:1928,carrington:1989}, the nuclei of tritium and $^3$He \cite{collard:1965,cruz:2020}, to the helium trimer \cite{schollkopf:1996,kunitski:2015}. The three-body problem is of fundamental theoretical interest, since it displays phenomena which are markedly different from the physics of just two particles. For pure contact interactions, Thomas predicted a collapse \cite{thomas:1935}, as the effective potential, an inverse square law in the hyperradius, supports infinitely deep bound states.
Efimov studied particles in the regime of large two-body scattering lengths, and found a universal ``staircase'' of trimer bound states, characterized by a constant ratio between successive energy levels \cite{efimov:1970}.

Efimov's scenario, originally proposed for nuclear physics, was studied in detail both experimentally and theoretically in the field of ultracold gases \cite{naidon:2017}. Efimov states were first observed in a gas of identical bosonic cesium \cite{kraemer:2006}, and later several successive states were demonstrated~\cite{zaccanti:2009,huang:2014,tung:2014,pires:2014}. The majority of observations were via particle loss at particular values of the scattering length. Exceptions are studies of three distinguishbable fermions of $^6$Li, where Efimov states were directly observed in radiofrequency (rf) spectra \cite{lompe:2010,nakajima:2011}, allowing the determination of binding energies.

The properties of the Efimov trimers greatly depend on the relative masses of the involved particles, especially when not all mutual two-body interactions are large \cite{naidon:2017}. For one light atom interacting with two heavy ones the Efimov effect becomes more prominent: successive energy levels are more closely spaced and easier to observe \cite{tung:2014,pires:2014}. For the opposite
 heavy-light-light combination which we study here, states with the true Efimov universality only appear exceedingly close to unitarity and at vanishing binding energies. 
 However, we show that this does not preclude the formation of a single weakly bound ``halo" trimer \cite{nielsen:1998}, which still retains some universal properties.

\begin{figure}[b!]
    \centering
    \includegraphics[width=0.95\columnwidth]{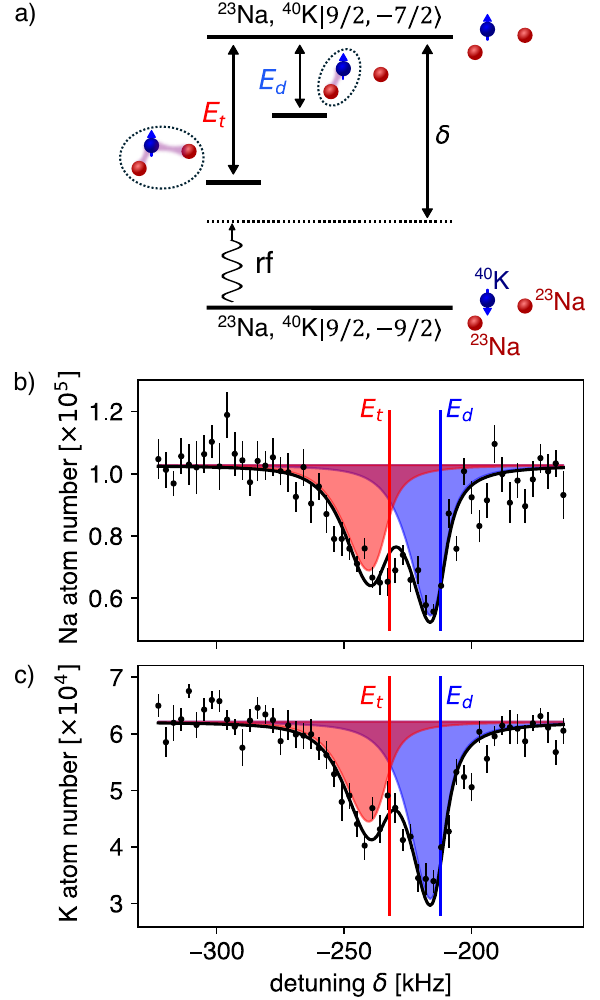}
    \caption{Detection of halo trimers using radiofrequency spectroscopy. (a) Schematic representation of free-to-bound transitions into Na-K dimer and Na$_2$-K trimer states with binding energies $E_d, E_t$, respectively, red-detuned from the bare atomic hyperfine transition. (b-c) Depletion spectra taken at 101.8 G, where losses due to resonant coupling to the dimer state and trimer state are present in (b) Na and (c) K. The underlying dimer and trimer fitted lineshapes are shown as blue and red shaded regions, respectively, with solid lines marking the binding energies and solid curves showing the summed lineshape.}
    \label{fig:intro}
\end{figure}

In this work, we directly observe such a halo trimer bound state using rf-spectroscopy as illustrated in Fig.~\ref{fig:intro}. The trimer is formed by two light $^{23}$Na-bosons surrounding a heavy ``impurity'' atom, fermionic $^{40}$K. The fermion interacts resonantly with the bosons near a Feshbach resonance, with scattering length $a$, while the bosons weakly repel each other (Bose-Bose scattering length $a_{\rm BB} = 56.5 a_0$, based on \cite{knoop:2011}).

We may understand the existence of this weakly bound trimer state by first considering the extreme limit of an infinitely heavy impurity. For positive scattering length, a dimer bound-state exists at energy $E_{d}$. In the absence of interactions between the light bosons, the impurity simply acts as an external potential in which both bosons can occupy the same energy level, giving this trimer a binding energy of $2 E_{d}$. When bosons instead repel at short range, they cannot simultaneously occupy the bound state and the trimer energy is suppressed. Universal properties of trimers and larger clusters in this scenario have been studied in Refs.~\cite{shi:2018,yoshida:2018} for an effective three-body repulsion, finding that a universal trimer state always persists at least for near-unitary interactions. 

In our case, the fermionic $^{40}$K atom is only about twice heavier than the bosons, 
but in combination with the modest interboson repulsion, this mass imbalance is enough to yield a scenario qualitatively similar to the infinite mass case.  
The trimer bound state we observe has an energy close to $E_{d}$, implying that the second boson is, on average, found significantly further away from the heavy atom than the first. We can reproduce the measured binding energy curve of the trimer theoretically.

In section \ref{sec:exp} we layout the experimental methodology  and the experimental evidence for the observation of the trimer state. In Sec.~\ref{sec:trimprop} we discuss the properties of the trimer state, such as lineshape, energy and wave function, based on both experimental and theoretical analyses. After discussing the relevant loss processes and the relative dimer-trimer signal strengths (Sec.~\ref{sec:loss}), we will conclude our work in Sec.~\ref{sec:conclusion}.

\section{Experimental observations}\label{sec:exp}
We prepare a Bose-Fermi mixture of $^{23}$Na and $^{40}$K, trapped in an optical dipole trap with Na trapping frequencies ${2\pi \times} (18, 118, 125)$ Hz in their respective hyperfine ground states $\ket{F=1, m_F=1}_\mathrm{Na}$ and $\ket{F=9/2, m_F=-9/2}_\mathrm K \equiv \ket{\downarrow}_\mathrm K$, with typical initial peak $^{23}$Na densities of 0.4 -- 0.88 $\mu\mathrm{m}^{-3}$ and $^{40}$K densities of $1.1$ -- $2.4$ $\mu$m$^{-3}$. To probe the few-body bound states of interest minimizing many-body effects in our system, we prepare the mixture in the non-degenerate regime at
 $\approx$\,200\,nK. Thus for our densities we have $T/T_C \sim 1.4 - 1.8$ and $T/T_F \sim 0.7 - 0.9$, where $T_C$ is the critical temperature for Bose-Einstein condensation and $T_F$ is the Fermi temperature. The ratio of atom numbers is comparable so that we are able to observe the creation and subsequent loss of molecules in the population depletion of both species. 

To measure the binding energies of the NaK dimer and Na$_2$K trimer states, we perform rf association spectroscopy by driving a free-to-bound hyperfine transition of K, from $\ket{\downarrow}_\mathrm K$ to $\ket{\uparrow}_\mathrm K \equiv \ket{F = 9/2,m_F = -7/2}_\mathrm K$, in the presence of Na, as shown schematically in Fig.~\ref{fig:intro}a \cite{wu:2012}.  We work near a broad interspecies Feshbach resonance at $\sim$ 110G between $\ket{F=1, m_F=1}_\mathrm{Na}$ and the final state $\ket{\uparrow}_\mathrm K$ by applying magnetic fields from $100.5$ to $105$ G (see Fig.~\ref{fig:Feshbach}). This changes the scattering length from approximately $600$ to $1600 a_0$. This allows us to measure the trimer binding energies as it tracks the dimer, over an order of magnitude. We then probe the transitions by an rf driving pulse \footnote{Blackman pulse, window-averaged Rabi frequency on the bare atomic transition $\Omega_\mathrm{rf} \sim 2\pi \times 5$ kHz}. 
Subsequently, we detect the populations of Na and the two spin states of K using resonant absorption imaging. For the rf pulse, we choose a variable duration between $3$ and $250$ ms, to keep loss fractions similar at the various fields. 
Strong losses in the initial state populations indicate the presence of resonant rf coupling to a bound state. 

\begin{figure}[b!]
    \centering
    \includegraphics[width=0.95\columnwidth]{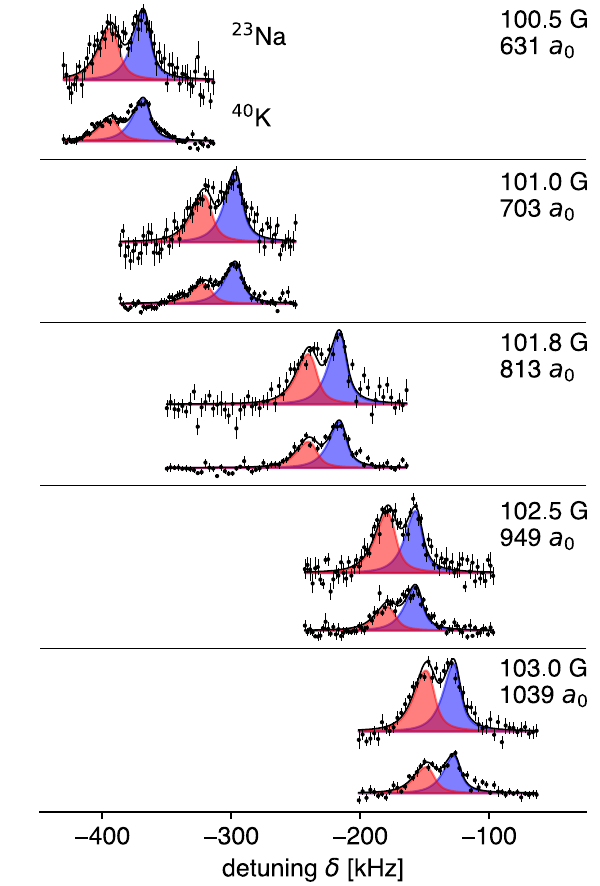}
    \caption{Rf-association of dimers and trimers. Depletion spectra of Na (K), shown as the upper (lower) trace within each panel with square (circle) markers, taken at B-fields between 100.5 and 103 G, in proximity to the Feshbach resonance at 110 G. The trimer feature (red shaded) closely tracks the dimer (blue shaded) at each field. The initially prepared K : Na ratios prior to the rf pulse vary between datasets from $0.3-0.65$.}
    \label{fig:waterfall}
\end{figure}

The initial interspecies interaction between Na and K in the initial hyperfine state $\ket{\downarrow}$ is strongly attractive, ranging from  from $-1400$ to $-1100 a_0$.
This attraction enhances two-body correlations and therefore wave function overlap to the bound states of interest, facilitating their observation.

\begin{figure}[t!]
    \centering
    \includegraphics[width=0.95\columnwidth]{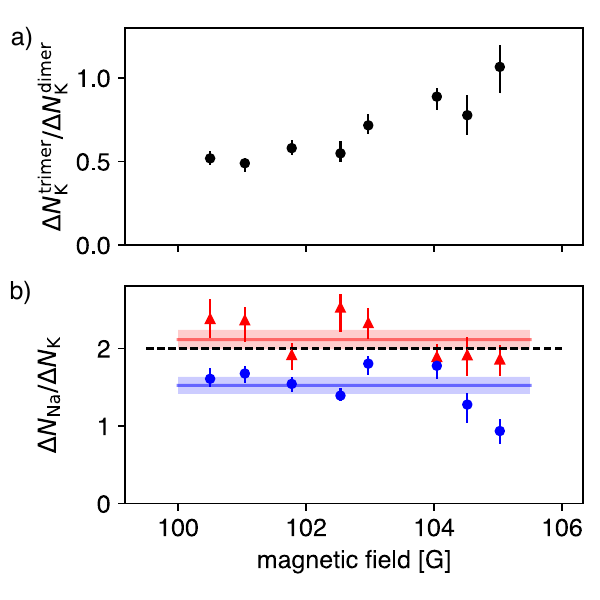}
    \caption{(a) The ratio between the fitted trimer and dimer feature heights, in the K spectra. (b) The intraspecies ratio of the fitted trimer (red triangles) and dimer (blue circles) peak depletions vs applied magnetic field. The trimer data clusters around $2:1$ Na-K loss ratio, whereas the dimer data lies between $1:1$ and $2:1$.  The solid lines (shaded regions) show the weighted mean and standard deviation of the estimated ratios, across magnetic fields.}
    \label{fig:lossratios}
\end{figure}

A compilation of rf depletion spectra for varying magnetic field is shown in Fig.~\ref{fig:waterfall}, where from top to bottom the Feshbach resonance is approached. In all of the spectra, the presence of the trimer state is clearly established by the double-peak structure of the rf-association feature. The trimer feature has comparable weight to the dimer peak for the whole range of magnetic fields. The relative peak weights as a function of magnetic field are shown in Fig.~\ref{fig:lossratios}a). We find that the trimer weight substantially increases as the resonance is approached. This can be understood from the decreasing relative binding energy leading to a larger wave function extent and therefore Franck-Condon factor, see Sec.~\ref{sec:trimprop}.

To fully understand our spectra, and to unambiguously establish the different nature of the dimer and trimer peaks, we need to consider not only the rf association processes but also possible loss processes. 
Indeed, without loss of the dimer and trimer populations, the signal detected in the Na imaging channel should remain unchanged. This is because our Na imaging probe is equally sensitive to the initial atom population as to the dimer and trimer populations, since the binding energies of dimer and trimer states ($\sim$100s of kHz) are small relative to the linewidth ($\sim$ 10 MHz) of the Na optical transition used for imaging. This allows the near-resonant Na imaging light to scatter and dissociate the molecules. 

The observed depletion of the Na atoms can be attributed to three-body loss processes: the relaxation of Feshbach dimers into a deeper rovibrational state upon collision or interaction with a third atom. The excess energy is released as kinetic energy, leading to the ejection of both the atom and the molecule from the trap \cite{wolf:2017}. For trimers this process can happen internally, leading to relaxation into a deeply bound dimer and free atom (Na$_2$K $\rightarrow$ NaK + Na). This is consistent with our observation that the association and decay of trimers depletes the Na and K initial populations in a $2:1$ ratio, as shown in Fig.~\ref{fig:lossratios}b). In contrast, the dimers can be lost through dimer-atom collisions with free atoms of either species (NaK-K and NaK-Na collisions). The presence of these two loss mechanisms, with $2:1$ and $1:2$ ratios for Na to K depletion, respectively, is also consistent with the observed depletion ratio of less than $2:1$ for the dimer feature. 

\begin{figure}[b!]
    \centering
    \includegraphics[width=0.95\columnwidth]{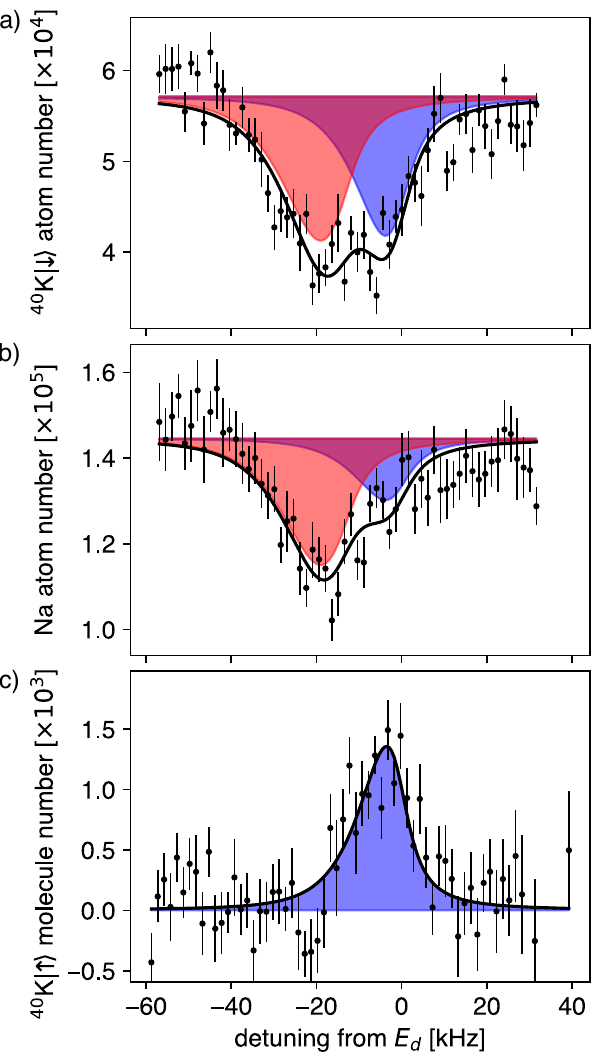}
    \caption{(a-b) Depletion spectra taken with a 30 ms Blackman-envelope rf pulse, at $B=105$ G ($a\approx 1633 a_0$). The loss in Na population on the dimer feature is strongly suppressed here. (c) A spectrum taken by applying a shorter 3 ms square pulse, with pulse parameters chosen to observe the arriving dimer and trimer populations in the final state $\ket{\downarrow}_\mathrm{K}$. There is only one spectral feature here, with the trimer arrivals being noticeably absent, due to their short intrinsic lifetime relative to the rf pulse duration.}
    \label{fig:arrivals}
\end{figure}

The difference in the loss dynamics between dimers and trimers is most striking near the Feshbach resonance, as shown  for $B=105$ G in Fig.~\ref{fig:arrivals} (scattering length $\approx 1600 a_0$). Here we see both the dimer and trimer features with roughly equal weights in the K depletion, whereas the Na depletion is strongly suppressed near the dimer resonance, indicating a lower rate of dimer-Na collisional loss. 

A consequence of the longer dimer lifetime is that near the Feshbach resonance we can directly detect the surviving dimers, as shown in Fig.~\ref{fig:arrivals}c. For this, the rf pulse duration is chosen comparable to the dimer decay time. Our measurement of the surviving dimer population after applying the rf pulse allows us to bound its $1/e$ decay time at $\gtrapprox 1.4$ ms at this magnetic field. 

We cannot detect any arrival signal on the trimer feature within our accessible experimental parameter space and our signal-to-noise (SNR), indicating a significantly shorter trimer lifetime. Given the length of the pulse and the SNR, we can upper-bound the trimer lifetime to $\sim$1 ms. By attributing the entire convolution kernel width used in the fitting functions to lifetime broadening, we can provide a lower bound of $\sim$20 $\mu$s. This is likely to be an underestimate of the lifetime however, since there are known factors such as technical noise that also contribute to broadening.

In conclusion, the difference in lifetimes close to the resonance, the different binding energies and the Na/K loss ratios are three physical features that allow us to clearly distinguish between the products, dimer and trimer, created with the rf drive.

\section{Trimer properties} \label{sec:trimprop}

\subsection{Lineshapes}

Having established the existence of the halo trimer experimentally, we move on to study its properties with the help of a theoretical analysis.
To obtain the dimer and trimer binding energies from the data, a good knowledge of the dimer and trimer lineshapes is required. In previous works on the $^{6}$Li trimer, the trimer lineshapes were fitted using phenomenological fit functions such as Lorentzian or Gaussian lineshapes \cite{lompe:2010,nakajima:2011}. Here we attempt to determine the energy more accurately by employing lineshapes based on theoretical modeling.
The total lineshape of the rf-association feature can be modeled as the
 weighted sum of the dimer and trimer lines. Both lines are intrinsically asymmetric due to the finite temperature and the free-to-bound nature of the rf-transition. 
 
 In a thermal gas, the dimer lineshape in the universal limit is well known \cite{chin:2005} and given by (setting $\hbar=1$)
\begin{equation} \label{eq:lsdim}
    F_d(\omega)\sim \frac{2}{\pi}\frac{k \kappa_d (1- \kappa_d a_i)^2}{(k^2+\kappa_d^2)^2(a_i^2 k^2+1)} \exp \left(-\frac{E}{k_B T} \right),
\end{equation}
where $\omega$ is the transition frequency and $\hbar \omega = E_d - E$. We further defined the initial scattering length $a_i$, temperature $T$, the initial state relative momentum $k=\sqrt{2 \mu_{\mathrm{NaK}} E}$ (with the Na-K reduced mass $\mu_{\mathrm{NaK}}$) and the dimer momentum scale $\kappa_d=\sqrt{2 \mu_{\mathrm{NaK}} |E_d|}$

The trimer lineshape cannot be captured in a simple analytical form, but the low-energy threshold behavior can be deduced straightforwardly.
We first briefly revisit the dimer case. The two-particle s-wave scattering wave function $f_d(k,r)$ is given by 
\begin{equation}
f_d(k,r) = \sqrt{\frac{2\mu_{\mathrm{NaK}}}{\pi k}} \sin(kr+\delta_k),
\end{equation}
where $r$ is the interparticle distance and 
$\delta_k$ is the phase shift, which for short-range interactions is given by
\begin{equation}
\lim_{k\rightarrow0} k \cot(\delta_k) = -\frac{1}{a}+\frac{r_{\mathrm{eff}} k^2}{2} + \mathcal{O}(k^4).
\end{equation}
for effective range $r_{\mathrm{eff}}$. Importantly, $\delta_k$ is linear in $k$ for small $k$. The Franck-Condon factor setting the lineshape is now given by the overlap of the scattering wave function $f_d$ with the localized bound-state wave function $\psi_{d}$. This yields
\begin{equation}
\lim_{k\rightarrow0} \left|\int_0^{\infty} dr \psi_{d}^{\ast}(r) f_d(k,r) \right|^2 \sim k \sim \sqrt{E},
\end{equation}
which we recognize from Eq.~\eqref{eq:lsdim}. In the experimentally relevant regime, the dimer is strongly bound compared to the initial thermal energy, so that the lineshape is well-approximated by $F_d \sim \sqrt{E} \exp(-E/k_BT)$. The momentum-dependence of the denominator in Eq.~\eqref{eq:lsdim} plays a minor role at the experimental temperature, except to set the prefactor of the expression to ensure the correct normalization of the wave function.

To consider the threshold scaling for the trimer wave function we employ the adiabatic hyperspherical approach (see appendix \ref{app:hyperspherical}). One can show that the three-body analogue of the s-wave scattering wave function is given by
\begin{equation} \label{eq:asymp_3body}
f_t(K,R) = \sqrt{\frac{2\mu R}{\hbar^2}}  [J_2(KR) - \tan{\Delta_K} Y_2(KR)],
\end{equation}
where $\mu$ is the three-body reduced mass, $R$ and $K$ are the hyperradius and the hyperradial momentum ($K=\sqrt{2 \mu E}$ for collision energy $E$), $J_2(KR)$ and $Y_2(KR)$ are regular and irregular spherical Bessel functions of second order and $\Delta_K$ is the three-body phase shift. In the three-body case the threshold law gives that $\tan (\Delta_K) \sim K^4$. Hence, similar to the dimer case, both terms in Eq.~\ref{eq:asymp_3body} have the same scaling, so that for $KR\rightarrow 0$, $f_t(K,R)\sim K^2$. Making a similar argument as in the dimer case, we can show that the wave function overlap between three-body scattering states and the localized trimer wave function $\psi_t$ scales as 
\begin{equation}
\lim_{K\rightarrow0} \left|\int_0^{\infty} dR \psi_{t}^{\ast}(R) f_t(K,R) \right|^2 \sim K^4 \sim E^2.
\end{equation}
The threshold behavior is therefore markedly different from the dimer case, leading to a larger temperature shift of the trimer peak than the dimer peak (in the low temperature limit). The different energy dependence of the trimer lineshape can be understood from a phase-space argument: for the low-energy contributions to the lineshape all the particles in the bound state need to be simultaneously at a small momentum, which is much more unlikely for the trimer than the dimer. 

Whereas the dimer is strongly bound compared to the thermal energy scale, this is no longer true for the halo trimer. Here the binding energy of the third atom to the dimer is comparable to the thermal energy. Hence, it is important to consider the leading-order corrections to the threshold scaling. Tscherbul and Rittenhouse \cite{tscherbul:2011} derive a trimer lineshape for Efimov states close to the atom-dimer threshold, and for $Ka\ll1$:
\begin{equation} \label{eq:trimer_lineshape}
F_t(\omega)\sim \frac{(2\kappa_t^2+K^2-2\kappa_t\sqrt{\kappa_t^2+K^2})^2}{K^4} \exp \left(-\frac{E}{k_B T} \right),
\end{equation}
where $\omega=E_t - E$, and $\kappa_t=\sqrt{2\mu(E_d-E_t)}$. Here the wave function consists of a dimer plus a weakly bound additional atom, and therefore we would expect a similar lineshape as in the case studied here. We indeed find a good agreement of this lineshape with the lineshape from hyperspherical calculations for the relevant low-energy regime. We do not account for the initial-state scattering length in the lineshape, because it is small compared to the thermal de Broglie wavelength. Therefore the lineshape is not much affected in the experimental regime. Note that this is different for the signal strength, which is substantially enhanced for initial-state attraction.

For the final fit function used to approximate the lineshape, we convolve the above-mentioned lineshape with a 8-10 kHz FWHM Lorentzian kernel and observe good agreement with the data, as can be viewed in Figs.~\ref{fig:intro}, \ref{fig:waterfall}, and \ref{fig:arrivals}.
The width of the experimental broadening is comparable to the intrinsic linewidth of the dimer and trimer features. The predominant source of broadening is given by rapid fluctuations of the applied magnetic field caused by technical noise. However, the atomic transition is observed to have a width of $\sim$5 kHz, and therefore the dimer and trimer peaks appear to experience additional broadening. We expect this additional broadening to be predominantly caused by the time-variation of the density profile and the temperature caused by the rf-association and loss. The trimer and dimer lines further experience many-body shifts and broadening, which we expect to be of the order of a kHz.

\begin{figure}[t!]
    \centering
    \includegraphics[width=0.95\columnwidth]{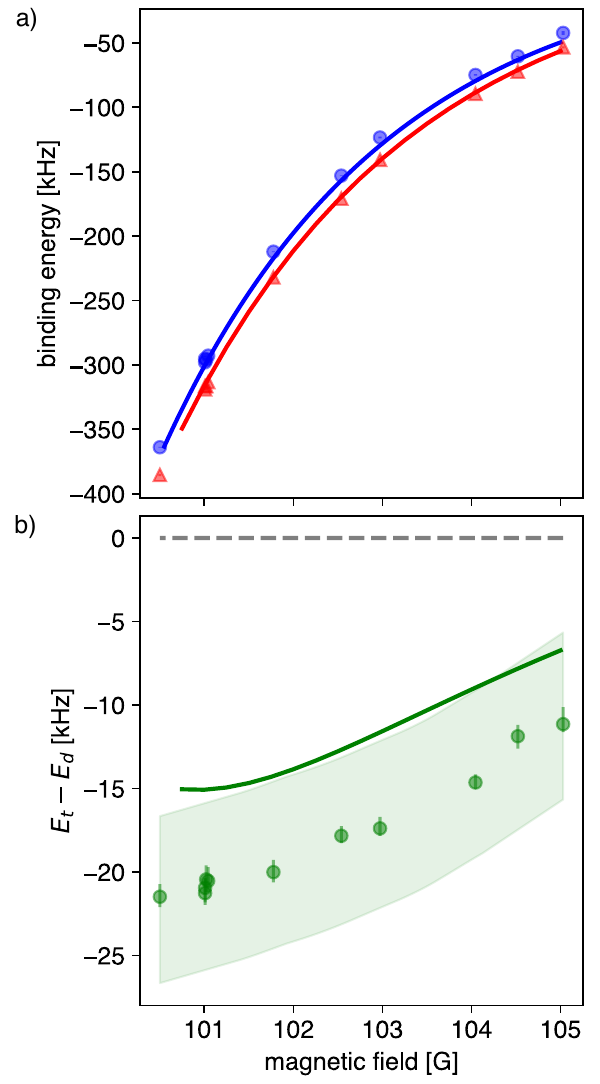}
    \caption{(a) The measured binding energies of the dimer (blue) and trimer (red) vs applied magnetic field. Error bars showing 68\% confidence intervals are found by bootstrapping and are within the markers. (b) Difference in measured binding energies of the dimer and trimer (green circles) along with the theoretical prediction (solid line). The shaded region shows the FWHM of the Lorentzian convolution kernel used to fit the lineshapes, setting a frequency scale for systematic error of the fit model.}
    \label{fig:waterfallSummary}
\end{figure}

\subsection{Trimer binding energy}
 
The fitted dimer and trimer binding energies are summarized in Fig.~\ref{fig:waterfallSummary}a. Their spacing, shown in Fig.~\ref{fig:waterfallSummary}b, is a small fraction (5-20\%) of the dimer binding energy across all our measurements. This strongly contrasts with the simplified fictitious case discussed in the introduction of a trimer state where an infinitely heavy impurity binds two non-interacting bosons, thus having a binding energy twice that of the dimer. The impurity having a finite mass only lowers the energy compared to the infinite-mass case due to the Efimov interaction mechanism \cite{naidon:2017,christianen:2022}, meaning that the suppression of the trimer energy is predominantly due to the interboson repulsion. 

Interestingly, the interboson repulsion does not have to be resonant to suppress the trimer binding energy. In our case the modest interboson repulsion of approximately $56.5 a_0$ already has this effect, even though the boson-impurity scattering length is more than an order of magnitude larger (ranging from $600-1600 a_0$ for experimental parameters). As discussed in Refs.~\cite{shi:2018,yoshida:2018} multichannel effects can also give rise to an additional effective repulsion, but this effect is strongest for narrow resonances. Here the same qualitative picture of the trimer state lying slightly below the dimer was predicted. More detailed investigations of multichannel effects for the current system, such as the presence of a virtual bound state and the interplay with the other nearby Feshbach resonances, are left for future work \cite{ahmed-braun:2024}.

We theoretically compute the dimer and trimer energies based on input from realistic Na-K and Na-Na interaction potentials  \cite{hartmann:2019,knoop:2011} (see Apps. \ref{app:hyperspherical} \ref{app:potentials} for more details). In short, we construct real-space model potentials using Gaussians, which match the scattering length and effective range of the realistic interaction potentials. All the parameters are chosen without fitting or matching to the experimental data.  We find that for the magnetic fields under consideration, the theoretical dimer energy is accurately represented by the universal expression
\begin{equation} \label{eq:Edim_effrange}
E_{d}=-\frac{1}{2 \mu_{\mathrm{NaK}} r_{\mathrm{eff}}^2}\left(1- \sqrt{1-\frac{r_{\mathrm{eff}}}{a}}\right)^2,
\end{equation}
where $a$ and $r_{\mathrm{eff}}$ both depend on $B$. The effective range varies approximately from $100-125 a_0$ over the experimental magnetic field range. This gives corrections to the dimer energy up to $20\%$ for $B=100.5 \ \rm{G}$ compared to the leading order term $(2\mu_{NaK}a^2)^{-1}$. The next order correction is about 1 kHz at this magnetic field, and much smaller closer to the resonance.

The theoretical dimer and trimer energies are plotted along the experimental data in Fig.~\ref{fig:waterfallSummary}a and  \ref{fig:waterfallSummary}b. From Fig.~\ref{fig:waterfallSummary}a we see that the theoretical dimer and trimer energies agree with the experimental results up to 5-10 kHz. However, in the difference plot we see that the theoretical dimer-trimer splitting is clearly smaller than the splitting obtained from the fits to the experimental data. Since the absolute difference is on the order of 5 kHz, which is smaller than the experimental linewidth, there is a variety of explanations for this difference.

In the lineshape model, we have not accounted for many-body energy shifts of the lines. Due to the shallow trimer state, the atom-dimer scattering length is large and positive, which leads to a shift of the dimer line to smaller binding energies. The trimer-atom scattering length is not known. However, if it is negative this will downshift the trimer line, and if it is positive there exist larger bound states below the trimer we do not resolve. In both cases this would shift the spectral weight to larger binding energies. Consequently, we expect these many-body effects to be responsible for an increase in the dimer-trimer splitting fitted from the experimental data on the kHz-level as compared to the theoretically computed (few-body) energy splitting.

Other effects, such as saturation, also have the tendency to increase the observed dimer-trimer splitting. Taking this into consideration, the theoretical results are consistent with the experiment.

\subsection{Trimer structure} \label{subsec:trimer_structure}
\begin{figure}[t]
    \centering
    \includegraphics[width=0.95\columnwidth]{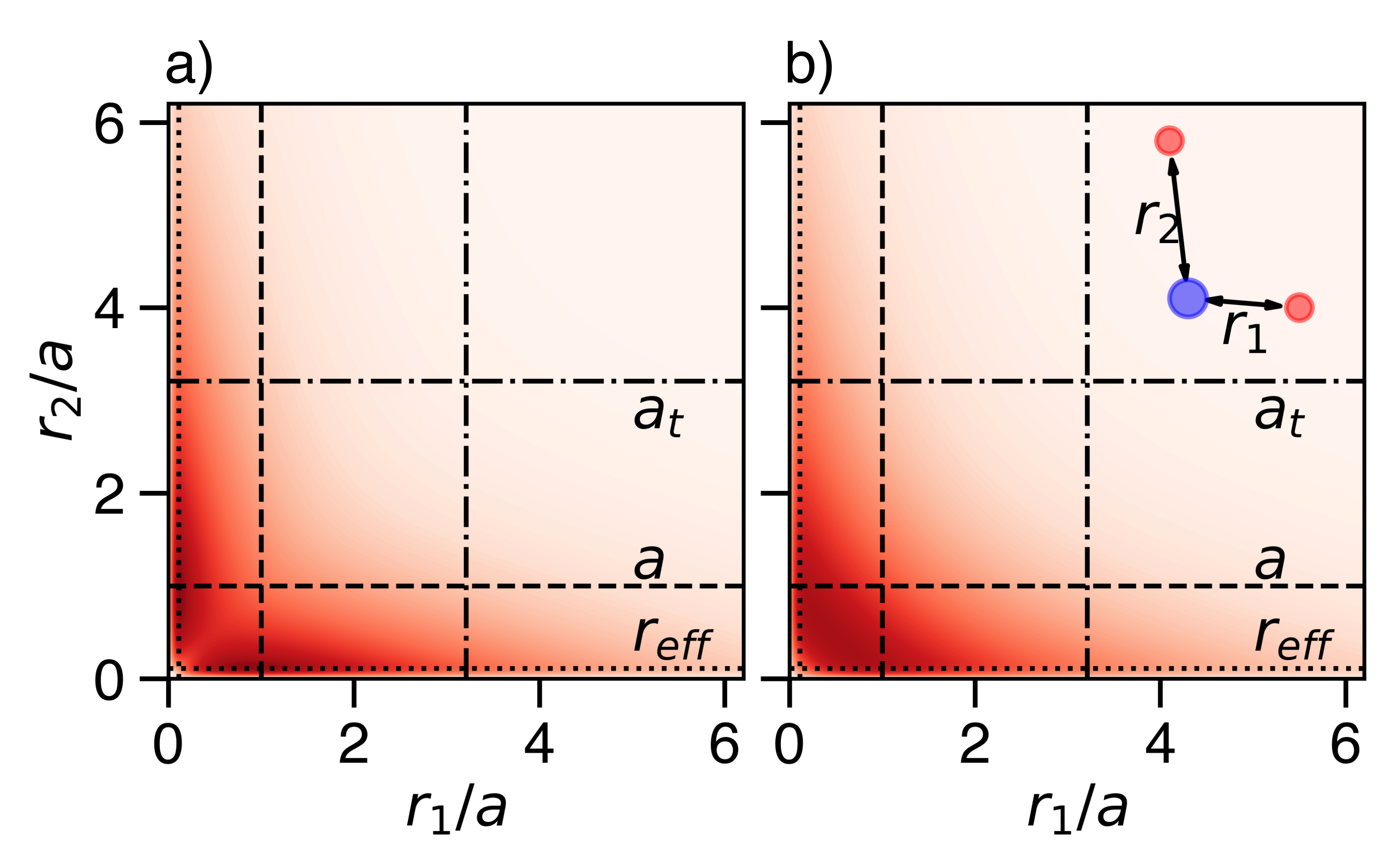}
    \caption{The trimer wave function amplitude as a function of the two Na-K distances $r_1/a$ and $r_2/a$ at $B=103$ G in units of the scattering length. a) shows the full numerical wave function, b) shows the result of the model from Eq.~\ref{eq:wftrim_simp}. The horizontal and vertical lines show the relevant length scales in the model, the Na-K scattering length $a$ (dashed), the trimer size $a_t=1/\kappa_t$ (dash-dotted) and the Na-K effective range $r_{\mathrm{eff}}$ (dotted).}
    \label{fig:trimerWavefunction}
\end{figure}

Since the theoretical trimer energy is consistent with the experiment, we can use the theoretical model to further analyze the internal structure of the trimer, as shown in Fig.~\ref{fig:trimerWavefunction}a). This figure shows the numerical trimer wave function for the magnetic field of 103 G as a function of the two Na-K distances $r_1$ and $r_2$. From the plot it is clear that the trimer can be viewed as consisting of a tightly bound dimer state, with a weakly bound additional third particle. 

Inspired by this result, we construct a model wave function of the form
\begin{equation} \label{eq:wftrim_simp}
    \psi_{t}(r_1,r_2)=\frac{1}{\sqrt{2N}}[ \psi_{d}(r_1) \psi_{ad}(r_2)+ \psi_{ad}(r_1) \psi_{d}(r_2)] .
\end{equation}
Here $N$ is a normalization factor, $\psi_{d}$ is the dimer wave function and $\psi_{ad}$ the atom-dimer wave function specifying the behavior of the third particle.
Both $\psi_d$ and $\psi_{ad}$ can be modeled using a similar functional form \footnote{This is the wave function solution for a separable potential of Yukawa form.}
\begin{equation} \label{eq:wftrim_simp2}
\psi_{d/ad}\sim \left[\exp(-\kappa_{d/t}r)-\exp \left(-\frac{r}{\rho_{d/t}} \right) \right],
\end{equation}
where
\begin{equation}
   \rho_d=\frac{3a}{8}-\frac{a}{8}\sqrt{9-16 \frac{r_{\mathrm{eff}}}{a}}.
\end{equation} 
and similarly for the $\rho_t$ in terms of the atom-dimer scattering length and effective range. The first term in Eq.~\eqref{eq:wftrim_simp2} is the asymptotic solution outside the range of the potential, whereas the second term captures a suppression of the wave function at short range. The precise form of this term is less important, as another term with a similar magnitude and extent would capture the physics equally well. Note that the $\kappa_t$ in this equation is defined using a different reduced mass than in Eq.~\eqref{eq:trimer_lineshape}, namely $\mu_{NaK}$ instead of the three-body reduced mass $\mu$. That is because here the spatial coordinate $r$ we are using corresponds to the boson-impurity distance instead of the hyperradius. The parameter $\rho_{d/t}$ characterizes the range at which the short-range wave function is suppressed. In particular, $\rho_d \approx r_{\mathrm{eff}}/3 $ for $r_{\mathrm{eff}} \ll a$.

We find that a wave function of the form of Eq.~\eqref{eq:wftrim_simp} has an overlap of around 98$\%$ with the full numerical trimer wave function for the whole range of magnetic fields probed in the experiment. The result from this construction for B=103 G is plotted in Fig.~\ref{fig:trimerWavefunction}b). For large values of either $r_1$ or $r_2$ the results in Figs.~\ref{fig:trimerWavefunction}a) and b) match very well. Since this is where most of the weight of the wave function is located, this leads to the large overlap which we find. However, as may be expected, the simplified wave function does not fully capture the short range details of the wave function for $r_{1},r_{2}<a$. In particular, the trimer wave function in a) shows a stronger suppression along the diagonal due to the interboson repulsion, which is not directly captured in b). 

The simplified wave function makes the interpretation of the trimer structure and its relevant length scales easier, since they can directly be read off from Eq.~\eqref{eq:wftrim_simp2}. These relevant length scales are indicated by the horizontal and vertical lines in Fig.~\ref{fig:trimerWavefunction}. The extent of $\psi_d$ is set by the dimer binding energy, where for B=103 G $1/\kappa_d\approx 1.06 a$. Similarly, the size of the trimer is set by $1/\kappa_t\equiv a_t$ (note that $a_t$ is not a scattering length). The peak of the dimer wave function is approximately set by the effective range $r_{\mathrm{eff}}$. From Fig.~\ref{fig:trimerWavefunction} we see that the peak of the trimer wave function lies around $a$, showing that the atom-dimer effective range is approximated well by the Na-K scattering length. We find this to be generally true over the magnetic field range which we study.

\section{Discussion: Loss processes and relative signal strengths} \label{sec:loss}

 \begin{figure*}[t!]
    \centering
\includegraphics[width=1.95\columnwidth]{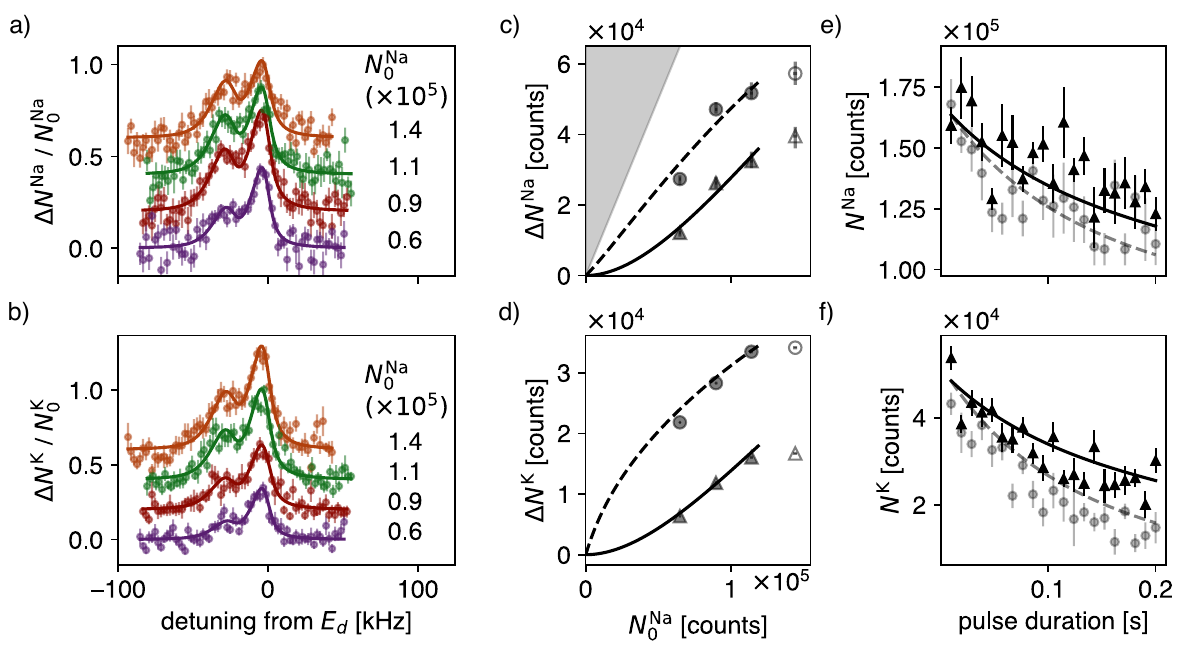}
    \caption{Dimer and trimer spectra as a function of the Na density. (a-b) Depletion spectra taken at 101 G, with initial corresponding peak Na density ranging from 0.4-0.88 $\mu$m$^{-3}$, shown as fractional loss of the initial population. Offsets have been added between spectra for visual clarity. (c-d) Peak heights of the fitted underlying dimer (circles) and trimer (triangles) features vs initial Na density. Solid (dashed) lines show fits to the two- and three-body collision models described in equation \ref{eq:trimerODE} (\ref{eq:dimerODE_K}, \ref{eq:dimerODE_Na}). The open-face markers represent the spectrum taken at the highest initial Na density, with a shorter rf pulse duration (100 ms, as opposed to 200 ms for the other spectra) to avoid unity depletion of the K population, and therefore deviate from the curve fits computed for 200 ms pulses. The grey shaded area in the upper plot has lower boundary corresponding to the maximum loss possible. (e-f) Initial state population depletion vs duration of rf pulse, with rf frequency chosen on the peak of the dimer (trimer) feature, corresponding to the circular (triangular) markers. Solid (dashed) lines show the prediction using the models described in (b).}
    \label{fig:density}
\end{figure*}

Since the formation processes of dimers and trimers require differing numbers of their constituent Na atoms, their association rates should depend differently on the density of Na atoms. We have measured this dependence and shown the results in the spectra in Fig.~\ref{fig:density}(a, b).

To model the trimer formation we use a simple rate equation with trimer formation coefficient $\Gamma_t$
\begin{equation}
\label{eq:trimerODE}
    \frac{dn_\mathrm K}{dt} = \frac{1}{2}\frac{dn_\mathrm{Na}}{dt} =  -\Gamma_{t} n_\mathrm{K}n_\mathrm{Na}^2.
\end{equation}
Here the densities of K, Na, are denoted by $n_\mathrm K, n_\mathrm{Na}$ respectively. 
We use that the trimer decay rate is much faster than the association rate in our experiment.

The dimer formation and decay processes are modeled with the relevant two-body processes, 
\begin{equation}
\label{eq:dimerODE_K}
    \frac{dn_\mathrm K}{dt} = -\Gamma_{d} n_\mathrm{K}n_\mathrm{Na} -\Gamma_{d, \mathrm{K}} n_dn_\mathrm{K},
\end{equation}
\begin{equation}
\label{eq:dimerODE_Na}
    \frac{dn_\mathrm {Na}}{dt} = -\Gamma_{d} n_\mathrm{K}n_\mathrm{Na} -\Gamma_{d, \mathrm{Na}} n_dn_\mathrm{Na},
\end{equation}
\begin{equation}
\label{eq:dimerODE_dimer}
    \frac{dn_d}{dt} = \Gamma_{d} n_\mathrm{K}n_\mathrm{Na} -\Gamma_{d, \mathrm{Na}} n_dn_\mathrm{Na} -\Gamma_{d, \mathrm{K}} n_dn_\mathrm{K},
\end{equation}
where $n_d$ denotes the density of dimers, $\Gamma_d$ is the density-normalized rf association rate, and $\Gamma_{d, \mathrm{\{species\}}}$ is the normalized loss rate due to atom-dimer collisions. Because $dn_d / dt \approx 0$ for most datasets in our experiment, we have a single timescale for population decay, set by the rate $\Gamma_{d} n_\mathrm{K}n_\mathrm{Na} \approx \Gamma_{d, \mathrm{Na}} n_dn_\mathrm{Na} + \Gamma_{d, \mathrm{K}} n_dn_\mathrm{K}$. We thus find that when fitting the rate coefficients, we are primarily sensitive to variations in $\Gamma_d$, which sets the timescale, and the ratio (but not the absolute values) of $\Gamma_{d, \mathrm{Na}} / \Gamma_{d, \mathrm{K}}$, which determines the ratio of population loss between species, as shown in Fig.~\ref{fig:lossratios}a.

Notice that for short rf pulse durations and small depletions $\Delta n / n \ll 1$, these two models predict the intuitive linear and quadratic Na density dependence we expect for the instantaneous formation rate of Na-K dimers and
 Na$_2$-K trimers, respectively. We find that each of these models captures the qualitative trends of the peak heights observed in experiment, with fitted coefficients $\Gamma_t / \Omega_{\mathrm{rf}}^2 = 9 \times 10^{-9}\,\mathrm{s} \cdot \mu\mathrm{m}^6$, $\Gamma_d / \Omega_{\mathrm{rf}}^2 = 1 \times 10^{-8}\, \mathrm{s} \cdot \mu\mathrm{m}^3$, and $\Gamma_{d, \mathrm{Na}} / \Gamma_{d, \mathrm{K}} = 2.5$, where $\Omega_\mathrm{rf}$ is the bare atomic Rabi frequency. This is shown in the fit lines of Fig.~\ref{fig:density}(c, d), as well as the time-dependence of population depletion with the rf pulse tuned to the dimer and trimer resonant features Fig. \ref{fig:density}(e, f). 

However, not all observed effects can be explained by this simple rate equation model. First, we observe in Fig.~\ref{fig:density}(c, d) that the results with a different pulse length deviate from the expected curve. 
Second, the relative magnitude of the trimer peak compared to the dimer peak as shown in  Fig.~\ref{fig:lossratios}(a) is larger than the theoretical predictions. Experimentally the dimer/trimer peak ratio varies from one to three. However, in a homogeneous system, the trimer signal is theoretically expected to be five to ten times lower than the dimer signal. 

Both of these effects can qualitatively be explained by saturation effects, heating, and trap dynamics. Especially since the peak K density is about three times higher than the peak Na density, the Na can be depleted in the center of the trap. Since the interspecies scattering length is large, the trap dynamics is expected to be diffusive. As a result, the repopulation of areas where the Na is depleted is slow, forming a limiting factor for the dimer and trimer association rates. The dimer creation rate is the most affected, since (theoretically) the dimer association rate is much higher than the trimer association rate. Heating will in addition have a strong effect on the density profiles of both species. We explore the effects of the trap dynamics and heating on rf association experiments in future work.

Further from the resonance than the experimentally studied regime, the trimer is expected to disappear into the atom-dimer continuum \cite{levinsen:2021}, leading to an atom-dimer scattering resonance \cite{knoop:2009}. In another experiment using the same Na-K mixture, but a different Feshbach resonance, indications of such a feature were observed \cite{chen:2022}.

\section{Conclusion and outlook} \label{sec:conclusion}
We have reported the first direct production of heteronuclear halo trimers, consisting of two light Na bosons bound to one heavy K-atom. From our rf-spectroscopy measurements we extracted the  trimer binding energy, lifetime, and the density dependence of the trimer formation rate. These features allow us to clearly distinguish the trimer peak in the rf-spectrum from the familiar Feshbach dimer signal.  Furthermore, we presented a theoretical analysis of the trimer energy and structure.

Reasonable technical improvements will allow the association of trimers within their decay time and thus direct measurements of their intrinsic lifetime, an active area of both experimental \cite{klauss:2017,yudkin:2019,yudkin:2024} and theoretical \cite{happ:2024} research. In addition, enhanced signal to noise will allow to more accurately measure the trimer energy, thus stringently testing theoretical models. This will help to advance our understanding of the universality of these trimers.

Our study is of direct relevance for several other research avenues in the domain of ultracold gases. The appearance of halo trimers should not be restricted to Na-K mixtures, because the conditions of having heavy ``impurities'' immersed in a Bose gas with modest positive scattering lengths are met in other mixtures as well.  When such mixtures are cooled into the degenerate regime, coupling to the trimer states is expected to lead to interesting higher-order correlations \cite{yoshida:2018b,levinsen:2021,christianen:2024}, which we start to explore in our future work \cite{robens:2024}. For the creation of bi-alkali molecules, the presence of weakly bound trimer states can have a detrimental effect. The commonly used magneto-association protocol for creating dimers by ramping across a Feshbach resonance \cite{duda:2023} may partially lead to the formation of short-lived trimers instead, thus reducing the efficiency. In particular, in the case of fermionic molecules, this leads to hole heating of the resulting molecular Fermi gas \cite{timmermans:2001}. Knowledge of the presence of the trimer state can therefore help to design better protocols to reach colder molecular samples.

\begin{acknowledgments}
We thank Richard Schmidt, Meera Parish, Jesper Levinsen, Servaas Kokkelmans and Jasper van de Kraats for useful discussions. A. Chuang acknowledges funding from the NSF GRFP. A. Christianen acknowledges funding from an ETH Fellowship. D. Ahmed-Braun acknowledges funding by the Research Foundation-Flanders via a postdoctoral fellowship (Grant No. 1222425N). 
This work was supported by the NSF through the Center for Ultracold Atoms and Grant PHY-2012110, AFOSR through grant FA9550-23-1-0402 and a MURI on Ultracold Molecules, an ARO DURIP and the Vannevar Bush Faculty Fellowship (ONR N00014-19-1-2631).
\end{acknowledgments}

\onecolumngrid
\appendix

\section{Theory}

\subsection{Adiabatic hyperspherical approach}\label{app:hyperspherical}

To model the trimer properties we employ the adiabatic hyperspherical approach. For an elaborate discussion, view for example Ref.~\cite{greene:2017}. Here we will give an overview of the main points. 

The first step in the construction is to define the mass-weighted Jacobi-coordinates
\begin{align}
\bm{\rho_1}&= \sqrt{\frac{\mu_{12}}{\mu}}(\bm{r_2}-\bm{r_1}), \\
\bm{\rho_2}&= \sqrt{\frac{\mu_{12,3}}{\mu}}\left(\bm{r_3}-\frac{m_1 \bm{r_1}+m_2 \bm{r_2}}{m_1+m_2} \right), 
\end{align}
where $\bm{r_1},\bm{r_2},\bm{r_3}$ are the positions of particles 1, 2 and 3 with masses $m_1, m_2, m_3$. We have further defined
\begin{align}
\mu&=\sqrt{\frac{m_1 m_2 m_3}{m_1+m_2+m_3}}, \\
\mu_{12}&=\frac{m_1m_2}{m_1+m_2}, \\
\mu_{12,3}&=\frac{(m_1+m_2)m_3}{m_1+m_2+m_3}.
\end{align}
We can now define the hyperradius $R=\sqrt{\rho_1^2+\rho_2^2}$, and its value is independent of the choice of the set of Jacobi coordinates. Together with two ``hyperangles" describing the internal configuration of the three-body system, and the three Euler angles describing its overall orientation, the three-body coordinate system is fully specified.

The Schr\"odinger equation can now be brought into the following hyperradial form
\begin{equation}\label{eq:SE_hyper}
    \left[-\frac{1}{2\mu} \frac{\partial^2}{\partial R^2} + \frac{15}{8 \mu R^2} + \frac{\hat{\Lambda}^2}{2\mu R^2}+V(R,\theta,\phi) \right] \psi = E \psi.
\end{equation}
Here $\hat{\Lambda}$ is the grand angular momentum operator, which describes the kinetic energy corresponding to the hyperangular motion and overall rotation for a given hyperradius. In the \textit{adiabatic} hyperspherical approach one solves this Schr\"odinger equation by first solving the hyperangular problem for every $R$, by diagonalizing $\frac{\hat{\Lambda}^2}{2\mu R^2}+V(R,\theta,\phi)$. We do this in practice by expanding the hyperangular wave function using an adaptive basis of B-splines.

After this procedure, the wave function $\psi$ is expanded into these hyperangular basis states for every $R$. The remaining hyperradial Schr\"odinger equation can then be solved by the $R$-matrix approach \cite{wang:2011}. By implementing the appropriate boundary conditions, one can both extract the three-body scattering properties and the bound-state properties.

Directly from Eq.\eqref{eq:SE_hyper} one can extract the form of the asymptotic solutions. Indeed, for the lowest hyperangular momentum state, and outside the range of the potential only the first two terms of the Schr\"odinger equation remain. The energy-normalized scattering solutions hence take the form of Bessel functions of the second order, as in Eq.~\eqref{eq:asymp_3body}.

\subsection{Interaction potentials and universality} \label{app:potentials}

Accurate interaction potentials for Na-K and Na-Na are available in the literature \cite{hartmann:2019,knoop:2011}. Directly employing these potentials including the full spin structure of the particles in our three-body calculations would be very challenging \cite{vandekraats:2024}. However, since the trimers found experimentally are very weakly bound we expect that model potentials having the same low-energy scattering properties as the realistic potentials should give similar trimer energies and wave functions. Since the Feshbach resonance under consideration is broad, we further assume that multichannel effects are not essential for the description of the trimer properties if the magnetic-field dependence of the scattering length and effective range is used as input for the calculations. A more thorough investigation of the multichannel effects on the trimer wave function is left for future work \cite{ahmed-braun:2024}.

\begin{figure}
    \centering
    \includegraphics[width=0.5\textwidth]{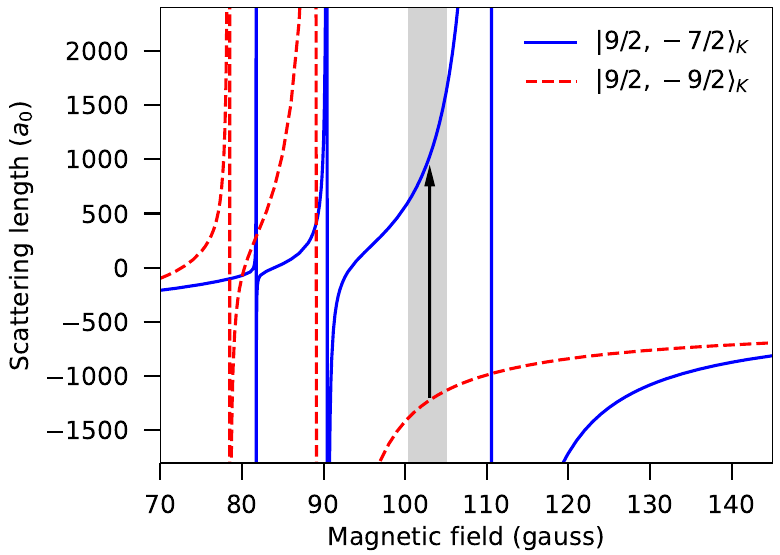}
    \caption{Scattering length (in $a_0$)  of scattering between $^{23}$Na
and $^{40}$K as a function of magnetic field. For $^{23}$Na the internal state is $|F = 1,m_F = 1 \rangle$,
for $^{40}$K the internal state is $|F = 9/2,m_F = -9/2\rangle$ (red dashed) and $|F = 9/2,m_F = -7/2\rangle$ (blue solid). The grey shaded area indicated the magnetic field ranged probed in the experiment. The curves are based on coupled-channels calculations with the Na-K interaction potentials from \cite{hartmann:2019}.}
    \label{fig:Feshbach}
\end{figure}

Concretely, we use coupled-channels calculations to compute the Na-K scattering length and effective range as a function of magnetic field (see Fig.~\ref{fig:Feshbach}). We find that the dimer energy in this magnetic field regime is described up to high accuracy using the universal formula \eqref{eq:Edim_effrange}. For every magnetic field, we now construct a Gaussian model potential with the same scattering length and effective range. For simplicity, we use a model potential without any deeper bound states in the potential aside from the dimer state under consideration.

The interboson scattering properties are also extracted using coupled-channels calculations, but they have negligible variation over the experimentally probed magnetic field range. Since the interboson scattering length is modest and positive, we model this with a repulsive potential without bound states. There does not exist a single repulsive Gaussian potential with the same scattering length and effective range as the realistic interboson potential. This can simply be remedied by choosing a sum of two Gaussian potentials. Now there is no unique choice, but we fix the potential by considering the phase shift over a larger energy regime. We constructed several different potentials with a similar energy-dependent phase shift, and found no substantial variation of the trimer-energy. More details and explicit parameters are given in Ref.~\cite{christianen_thesis:2023}.

\nocite{*}

\bibliography{halotrimer}

\end{document}